\documentclass[12pt,preprint]{aastex}
\usepackage{graphicx}
\newcommand{\SO}{XTE~J1650$-$500}
\newcommand{\LF}{low frequency}
\newcommand{\La}{$L_{1}$}
\newcommand{\Lb}{$L_{2}$}

\begin{document}


\title{{X}-ray temporal properties of {XTE J1650$-$500} during outburst decay }

\author{E. Kalemci\altaffilmark{1},
 J. A. Tomsick\altaffilmark{1},
 R. E. Rothschild\altaffilmark{1},
 K. Pottschmidt\altaffilmark{2,3,4},
 S. Corbel\altaffilmark{5},\\
 R. Wijnands\altaffilmark{6,7},
 J. M. Miller\altaffilmark{6},
 P. Kaaret\altaffilmark{8}
}

\altaffiltext{1}{Center for Astrophysics and Space Sciences, Code
0424, University of California at San Diego, La Jolla, CA,
92093-0424, USA}

\altaffiltext{2}{Institut f\"{u}r Astronomie und Astrophysik,
Abt. Astronomie, Sand 1, 72076 T\"ubingen, Germany}

\altaffiltext{3} {\emph{INTEGRAL} Science Data Center, Chemin d'\'Ecogia 16, 1290 Versoix, Switzerland}

\altaffiltext{4} {Max Planck Institut f\"{u}r Extraterrestische Physik, Postfach 1312, 85741 Garching, Germany}

\altaffiltext{5}{Universit\'e Paris VII and Service d'Astrophysique, CEA Saclay, F-91191 Gif sur Yvette, France}

\altaffiltext{6}{Center for Space Research, 70 Vassar Street, NE-80/6055, MIT, Cambridge, MA, 02139, USA}

\altaffiltext{7}{Chandra Fellow}

\altaffiltext{8}{Harvard-Smithsonian Center for Astrophysics, 60 Garden Street,
 Cambridge, CA, 02138, USA}


\begin{abstract}

     We investigated the temporal behavior of the new black hole transient 
\SO\ with the \emph{Rossi X-ray Timing Explorer (RXTE)} as the source made a 
transition to the low/hard state during the decay of the 2001 outburst. We 
find QPOs in the 4 -- 9 Hz range, enhanced time lags and reduced coherence 
during the state transition. We also observe a shift in the peak frequency of 
the noise component with energy during the transition. The evolution of the 
power spectrum as well as the lag and coherence behavior during the state 
transition are similar to the state transitions observed for other black hole 
sources, especially Cyg X-1. The temporal properties during the transition to 
the low state put constraints on the accretion geometry of \SO\ and may have 
implications for all black hole binary systems. We suggest a possible 
geometry and evolution of a jet+corona+disk system based on enhanced lags and 
peak frequency shift during the transition.
\end{abstract}



\section{Introduction}\label{sec:intro}
 
   Galactic black hole candidates (BHC) exhibit transitions between X-ray 
states distinguished by their different spectral and timing properties 
\citep{Tanaka95}. The classification of these states is not rigorously defined 
and is an active topic of debate. The accretion rate seems to be the most 
important parameter determining the spectral states, although it is argued by 
\cite{Homan01} that two independent parameters are needed to describe the 
states, and the second parameter could possibly be the size of the 
Comptonizing region. Five states have been frequently quoted: the off state, 
the low-hard state (LS), the intermediate state (IS), the high-soft state (HS) 
and the very-high state (VHS). They are named after their spectral and 
luminosity properties, but also show very distinct temporal and radio 
properties. The HS is dominated by a soft spectrum with little or no timing 
noise (rms amplitude less than a few \%), and the radio emission is quenched 
\citep{Fender99,Corbel00}. On the other hand, the LS is dominated by a power 
law component in the X-ray spectrum, which is often interpreted as being 
associated with a hot electron corona, and shows very strong band limited 
noise (up to 50\% rms amplitude) and quasi-periodic-oscillations (QPO). 
Moreover, the radio observations often reveal a compact jet during this state 
\citep[][and references therein]{Corbel01,Fender01}. The IS/VHS shows both an 
ultra-soft spectral component and a power-law tail with 1\% -- 15\% rms 
variability. Some of the BHCs show high frequency QPOs (with frequencies of 
several hundreds of Hz) in this state. The compact jet is also quenched in 
the IS/VHS, i.e. in any state where a soft, strong X-ray component exists 
\citep{Corbel01}. 

   The geometry of the accretion system in each state and the triggering 
mechanism for the state transitions are not well understood \citep{Nowak02}. 
 The most notable observational properties during the transition to the LS are
 (1) the appearance of QPOs and strong band limited noise in the power density 
spectrum (PSD) and the hardening of the power law component in the X-ray 
spectrum, possibly indicating the formation of a hot corona, (2) the evolution 
of the characteristic frequencies of the power spectrum which might indicate 
retreating of the inner-disk radius \citep{diMatteo99,Tomsick00,Kalemci01}, 
(3) optically thin synchrotron radio emission indicating plasma ejections and 
formation of compact radio jets \citep{Corbel00}, and (4) a decrease in 
coherence and an increase in time lags observed in Cyg X-1 which might be 
related to the radio jets \citep{Pottschmidt00}.

   Here, we describe a state transition found in the soft X-ray transient \SO\ 
in light of the observations above. \SO\ was first detected  by the 
All-Sky Monitor \citep[ASM;][]{Levine96} on board the \emph{Rossi X-ray Timing 
Explorer (RXTE)} in 2001 September \citep{Remillard_IAU01}. The optical and 
radio counterparts were identified by \cite{Castro_Tirado_IAU01} and 
\cite{Groot_IAU01}, respectively. \cite{Augusteijn_IAU01} suggest a long 
orbital period and a low-mass secondary. Pointed X-ray observations by 
\emph{RXTE} show strong \LF\ QPOs and band limited timing noise 
\citep{Wijnands_IAU01}, while \emph{Chandra} revealed ionized iron absorption 
lines \citep{Miller_ATEL02}. The X-ray flux started to decay 45 days after the
 beginning of the outburst, and a transition to the hard state occurred. 
During the hard state, additional radio measurements were made and an inverted 
spectrum indicative of a compact jet was observed (S. Corbel, private 
communication). Around 10 days after the state transition, the source was 
unobservable due to the Sun. After the Sun gap, the X-ray flux showed a 
modulation with a $\sim$14 day period \citep{Tomsick_iau02}. The mass function
 has yet to be measured, however both the X-ray spectra, and the temporal 
properties are similar to the known Galactic black hole candidates (BHC). 
Moreover, a recent \emph{XMM-Newton} analysis of the broad iron line feature 
suggests that \SO\ is a rapidly rotating black hole \citep{Miller02}.  

%
\begin{figure}
\plotone{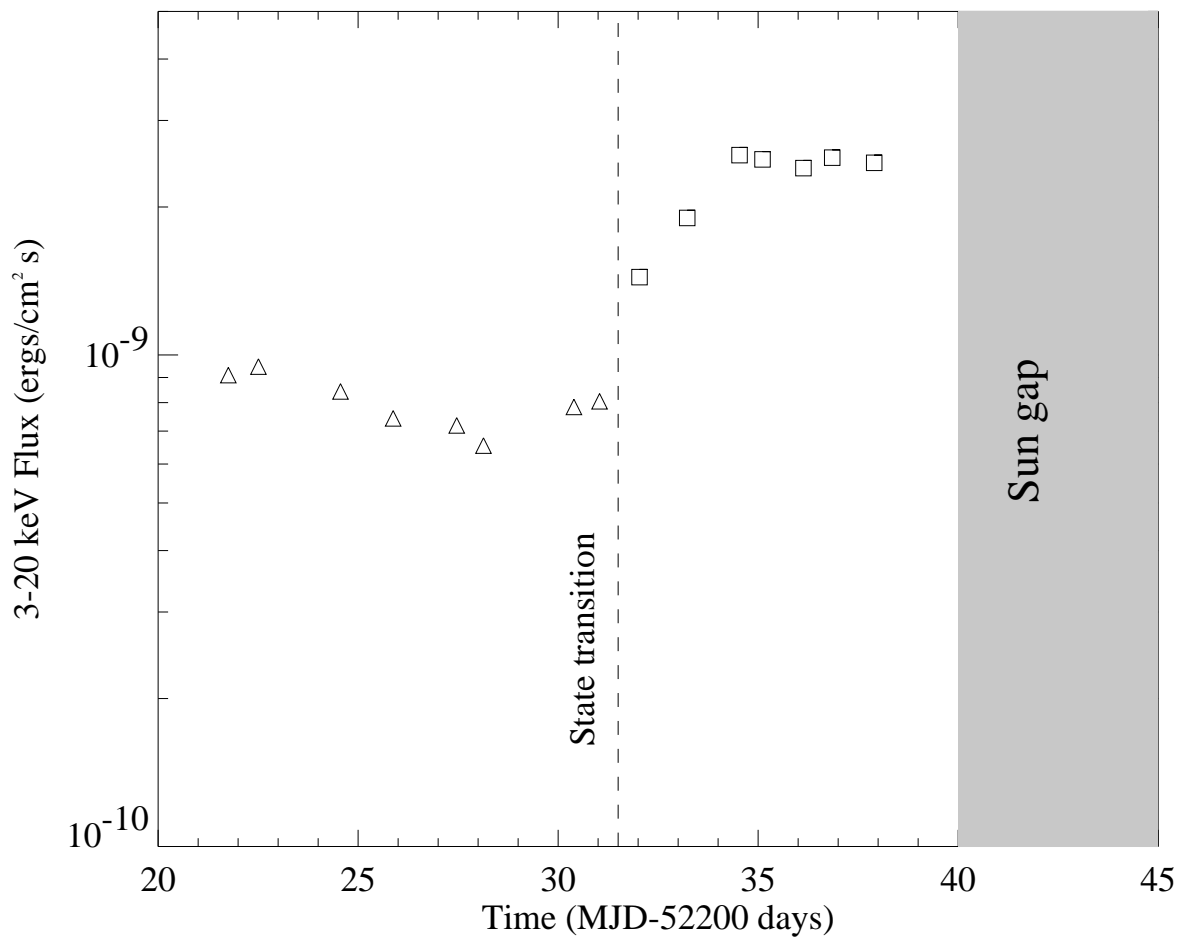}
\caption{\label{fig:lc} 
PCA light curve of \SO\ in 3 -- 20 keV band. The dashed line represents the 
time when the state transition occurred. The observations represented by 
squares are the ones that we focus on in this paper. The gray area is the 
time when the source is close to the sun and unobservable. The 3 -- 20 keV 
X-ray flux increases during the transition to the LS which is uncharacteristic 
for BHCs.
}
\end{figure}

  In this paper, we search for high frequency QPOs and study the evolution of 
the low frequency QPOs through the decay. We investigate the evolution of the 
continuum parameters of the power spectra during the transition, and compare 
the temporal properties to the previous observations and other BHCs. We also 
employ other temporal analysis tools related to the cross spectrum, namely the 
frequency dependent time lags and coherence function.


\section{Observations and Analysis}\label{sec:obs}

We analyzed the PCA/\emph{RXTE} (Proportional Counter Array, see 
\cite{Bradt93} for a description of \emph{RXTE}) data from observations of 
\SO\ from the start of the decay ($\sim$MJD\footnote{Modified Julian Date 
(MJD = JD - 2400000.5)}~52218) until the source was close to the sun between 
MJD~52240 -- 52266 (see Fig.~\ref{fig:lc}). When it was observable again, the 
flux had decayed significantly.  Until MJD~52230 the X-ray energy spectrum was 
typical of a HS black hole spectrum. Between MJD~52230 and 52232 (around the 
dashed line in Fig.~\ref{fig:lc}) the energy spectrum hardened.\footnote{The 
detailed spectral analysis of \SO\ during the outburst decay from the 
\emph{RXTE} and the \emph{Chandra} observations will be presented in a 
separate paper by \cite{Tomsick_inprep02}.} For the observation at MJD~52232, 
both the energy spectrum and the power spectrum showed significant changes and 
a transition to the hard state occurred. (The first square after the dashed 
line in Fig.~\ref{fig:lc}.) We call this observation Obs~1, and number the 
following observations according to their date of observation. Before the 
transition, the variability was very weak and the power spectrum was not above 
the noise level. We obtained a PSD by combining light curves from our 10 
observations in the HS and fitted it with a Lorentzian, which resulted in a 
95\% confidence rms amplitude upper limit of 4\%. The rms amplitude jumped to 
17\% for Obs~1, and a weak QPO at 8.71 Hz is observed for Obs~2. For the 
following observations until the sun gap, both the energy spectrum and the 
power spectrum showed characteristics of a typical hard state (see 
Fig.~\ref{fig:lor}). In this paper, we focus on the 7 observations after the 
state transition and before the sun gap, represented by the squares in 
Fig.~\ref{fig:lc}. Due to the drastic change in the flux and the low quality 
of the temporal data, the observations after the sun gap are not included in 
this work. 

   We used the standard \emph{RXTE} data analysis software FTOOLS 5.1 to 
extract light curves. For all observations, the data were accumulated in an 
event mode with $\rm 125\,\mu s$ time resolution and 64 energy channels. 
Typical integration times were 1 -- 2 ks per observation. We used all of 
the Proportional Counter Units (PCU) that were on simultaneously for the timing
analysis. The observations were made after the loss of the propane layer for 
PCU~0, and we dealt with the additional background as described in 
\cite{Kalemci01}. For each observation, we computed the power spectra and 
cross spectra using IDL programs developed at the University of T\"{u}bingen 
\citep{Pottschmidt02th} for three energy bands, 2 -- 6 keV, 6 -- 15 keV, and 
15 -- 30 keV and combined band of 2 -- 30 keV. Above 30 keV, the source is not
 significantly above the background. The PSD was normalized as described in 
\cite{Miyamoto89} and corrected for the dead time effects according to 
\cite{Zhang95} with a dead-time of $\rm 10\,\mu s$ per event. First, we 
searched for high frequency QPOs using short 1 second segments with a Nyquist 
frequency of 2048 Hz using the summed energy band of 2--30 keV. We did not 
detect any high frequency QPO in any of our observations. The 95\% confidence 
upper limits on the rms amplitude for QPOs above 50 Hz with a fixed quality 
value of $Q = 6$ are between 3.7 -- 4.4 \% for different observations. Next, 
using 128 second time segments, we investigated the low frequency QPOs and the 
timing properties of the continuum up to 256 Hz. We also analyzed the time 
lags and coherence function between 2 -- 6 keV and 6 -- 15 keV energy band 
light curves.
 

\section{Results}\label{sec:results}

   Historically, the PSD of BHCs during the low state has been modeled by a 
broken power law (or power laws with more than one break) plus narrow 
Lorentzians to fit the QPOs \citep{Nowak99,Tomsick00}. This model is a good 
representation of our PSDs also. However, recent papers successfully fit 
several BHC and neutron star PSDs with broad Lorentzians for the continuum and 
narrow Lorentzians for the QPOs \citep{vanStraaten01,Pottschmidt02,Belloni02}. 
When we fit our data with broad and narrow Lorentzians we see that this 
modeling represents the PSDs as well as or better than the broken power law in 
terms of $\chi^{2}$ statistics. Although we did both kinds of fitting, we only 
present the Lorentzian fits. This allows for an easier comparison to recent 
work, for example on Cyg X-1 by \cite{Pottschmidt02}.

\subsection{Lorentzian fits}

\begin{figure}
\plotone{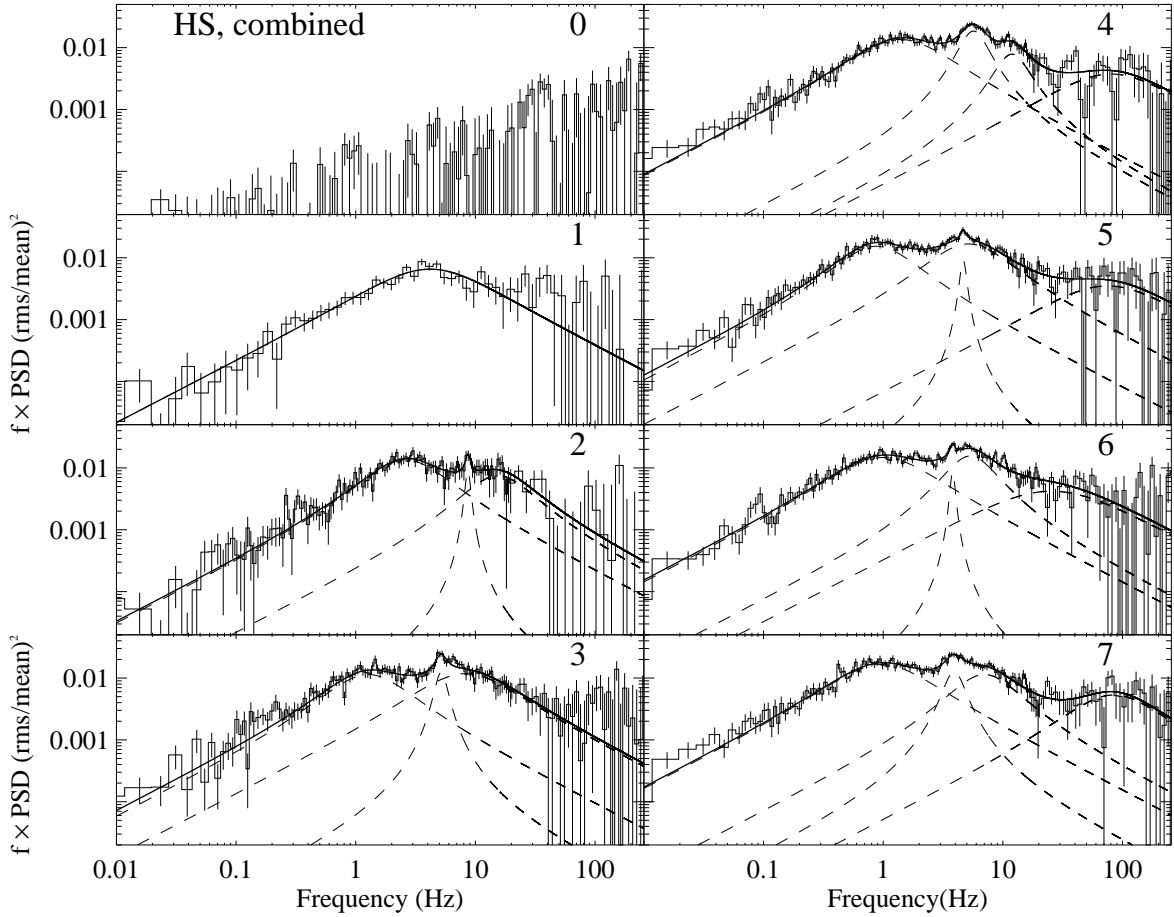}
\caption{\label{fig:lor} 
Power spectra and Lorentzian fits. The panel indicated by ``0'' is the power
spectrum from the combined light curves in the HS. Except for the HS power
spectrum, the solid line represents the overall fit and the dashed lines 
represent each component.
}
\end{figure}

We fit all our PSDs with Lorentzians of the form
\begin{equation}
L_{i}(f)\;=\;{{R_{i}^{2} \; \Delta_{i}}\over{2 \, \pi \; [(f-f_{i})^{2}+({1\over{2}}\,\Delta_{i})^2]}}
\end{equation}
where subscript $i$ denotes each Lorentzian component in the fit, $R_{i}$ is 
the rms amplitude of the Lorentzian, $\Delta_{i}$ is the 
full-width-half-maximum, and $f_{i}$ is the resonance frequency. A useful
quantity of the Lorentzian is the frequency at which its contribution to the 
total rms variability is maximum (hereafter peak frequency):
\begin{equation}
\nu_{i}\;=\;{f_{i}\;\left({{\Delta_{i}^{2}}\over{4 \, f_{i}^{2}}}+1\right)^{1/2}}
\end{equation}
In Fig.~\ref{fig:lor}, we plot the power spectra in the form of
PSD $\times$ frequency and over-plot the Lorentzian fits. In this figure, 
the Lorentzians peak at $\nu_{i}$ not at $f_{i}$. In this section, including 
the tables, Lorentzian frequency means the peak frequency, not the resonance 
frequency. Most of our observations contain a Lorentzian that is narrow (with
 $Q>2$, as compared to $Q<1$ for broad Lorentzians) which we call a QPO. The 
QPO frequencies reported here are the resonance frequencies to make them 
easier to be compared to previous reports. 

As seen in Fig.~\ref{fig:lor}, except for the first observation, each 
observation requires at least two broad Lorentzians to fit the continuum, and 
a QPO (see Table~\ref{table:par_lor} for the fit parameters). For Obs~1, the
95\% confidence upper limit on the rms amplitude of a QPO in 0.1 - 20 Hz range 
with a $Q = 6$ is 2.8 \%. The QPO is always the second peaked component in the
 PSD. Although the second peaked component in Obs~4 has a quality value of 
$\sim$1.5, it is still quoted as a QPO for consistency. Since they are narrow, 
the quoted resonance frequencies are very close to the peak 
frequencies\footnote{The peak frequency and the resonance frequency differ by 
5\% for the worst case of Obs~4.}. We call the broad Lorentzians in the fit 
$L_{i}$, where $i$=1 is the lowest frequency peak Lorentzian. The QPO is not 
counted and always referred as the QPO throughout the paper. 

\begin{figure}
\plotone{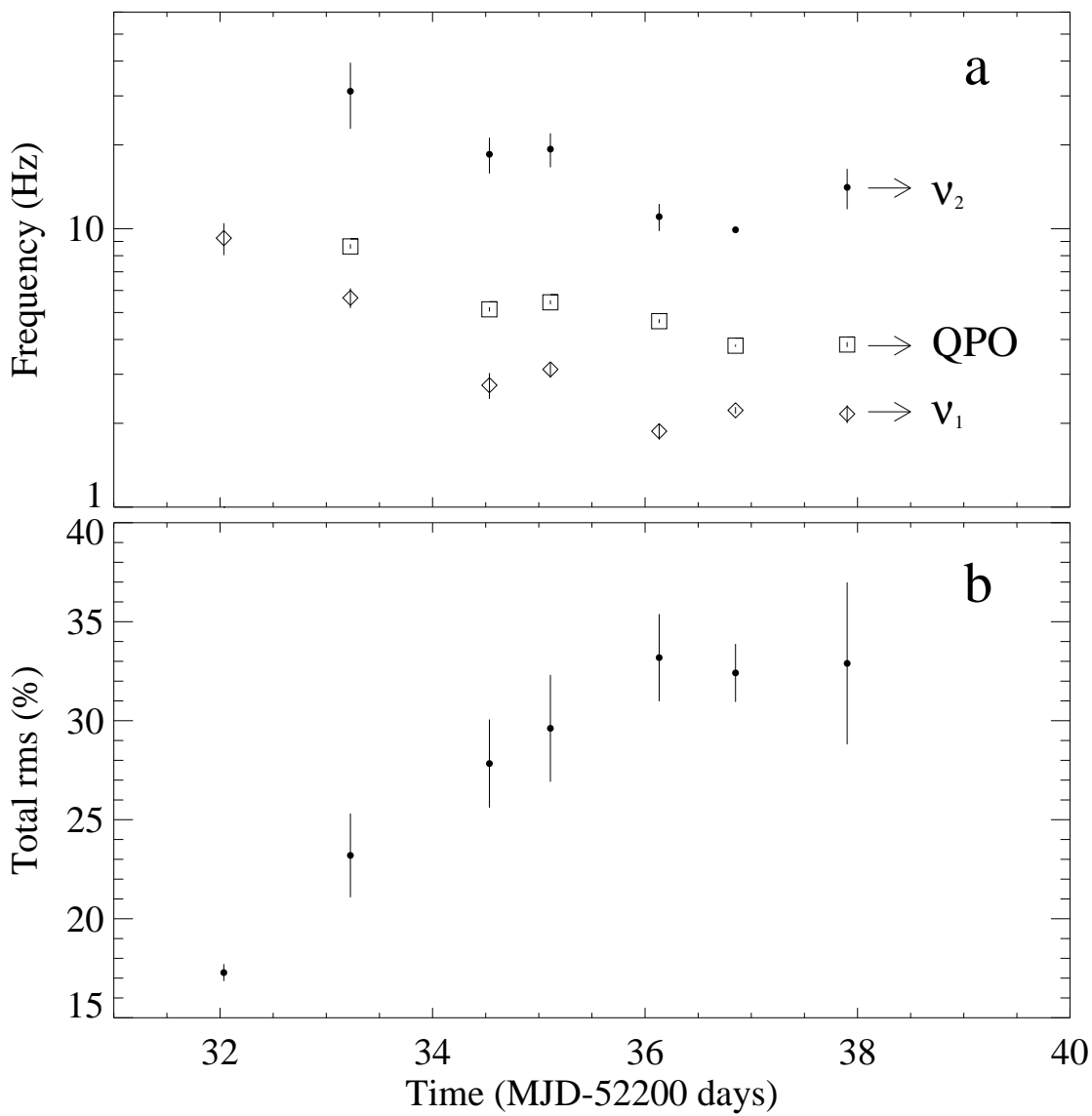}
\caption{\label{fig:par_lz} 
Lorentzian fit parameters: (a) Peak frequencies of the Lorentzians and the 
resonance frequencies of the QPOs. Filled circle represents $\nu_{2}$, open 
square represents QPO frequency and diamond represents $\nu_{1}$. (b) Total 
rms amplitude of the Lorentzians, including the QPO. During the HS, the total 
rms amplitude has 95\% confidence upper limit of 4\%.
}
\end{figure}

There is an overall shift to lower frequencies with time for both the wide 
Lorentzians and the QPOs (see Fig.~\ref{fig:par_lz}a). The overall rms 
amplitude increases with time, and settles around 35\% after Obs~5 as seen in 
Fig.~\ref{fig:par_lz}b. The rms amplitudes of individual Lorentzian components
do not show a regular trend and can be found in Table~\ref{table:par_lor}.

\begin{figure}
\plotone{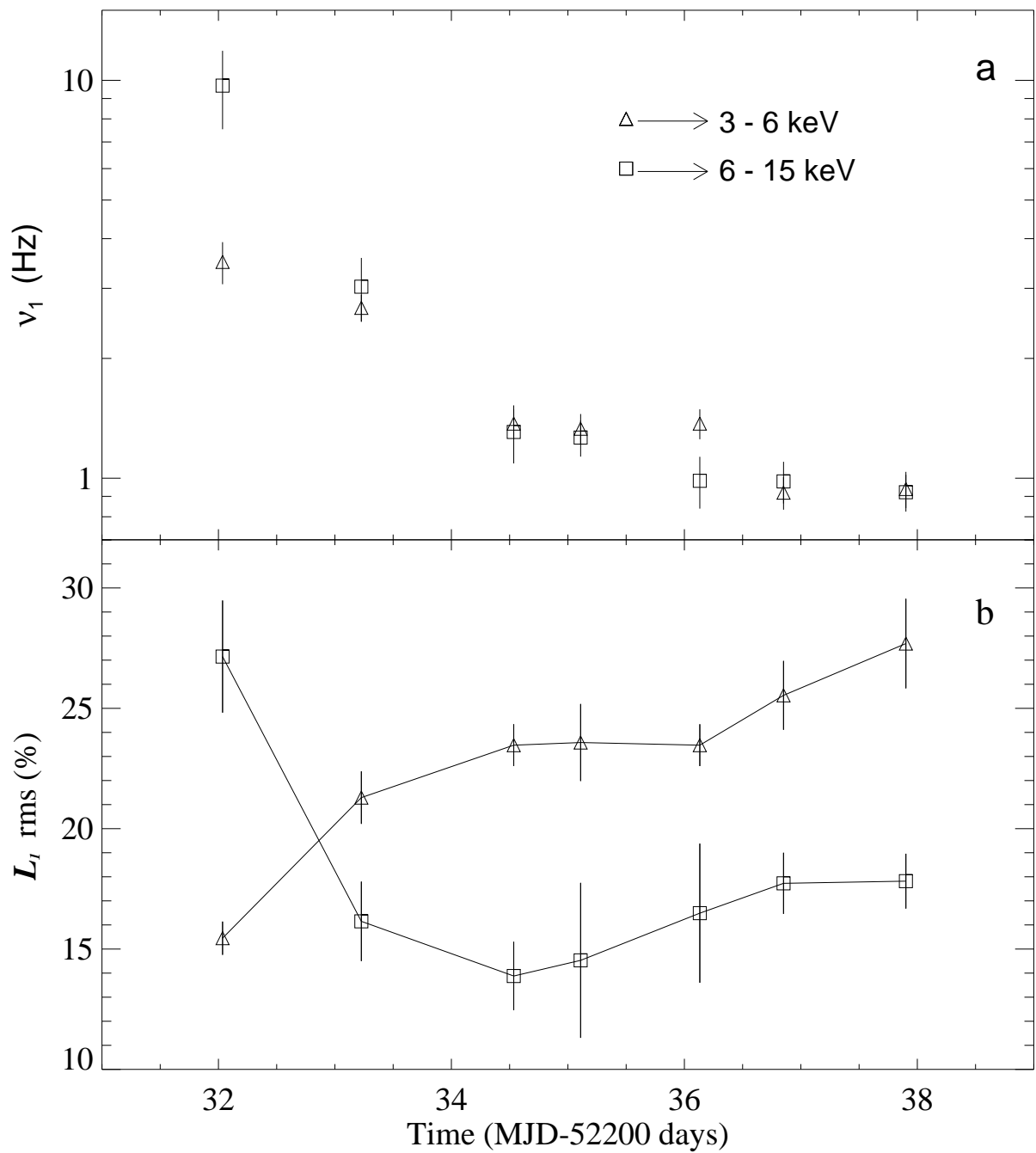}
\caption{\label{fig:frqsh} 
Energy dependence of PSD fit parameters: (a) Peak frequencies of \La . Squares 
represent the 6 -- 15 keV results, and triangles represent the 2 -- 6 keV 
results. The 6 -- 15 keV peak frequency for the first observation is clearly 
higher than the 2 -- 6 keV one. For the remaining observations, the peak 
frequencies are the same for both energy bands within 1-$\sigma$ error. (b) 
Rms amplitude in the two energy bands mentioned. 
}
\end{figure}

\begin{figure}
\plotone{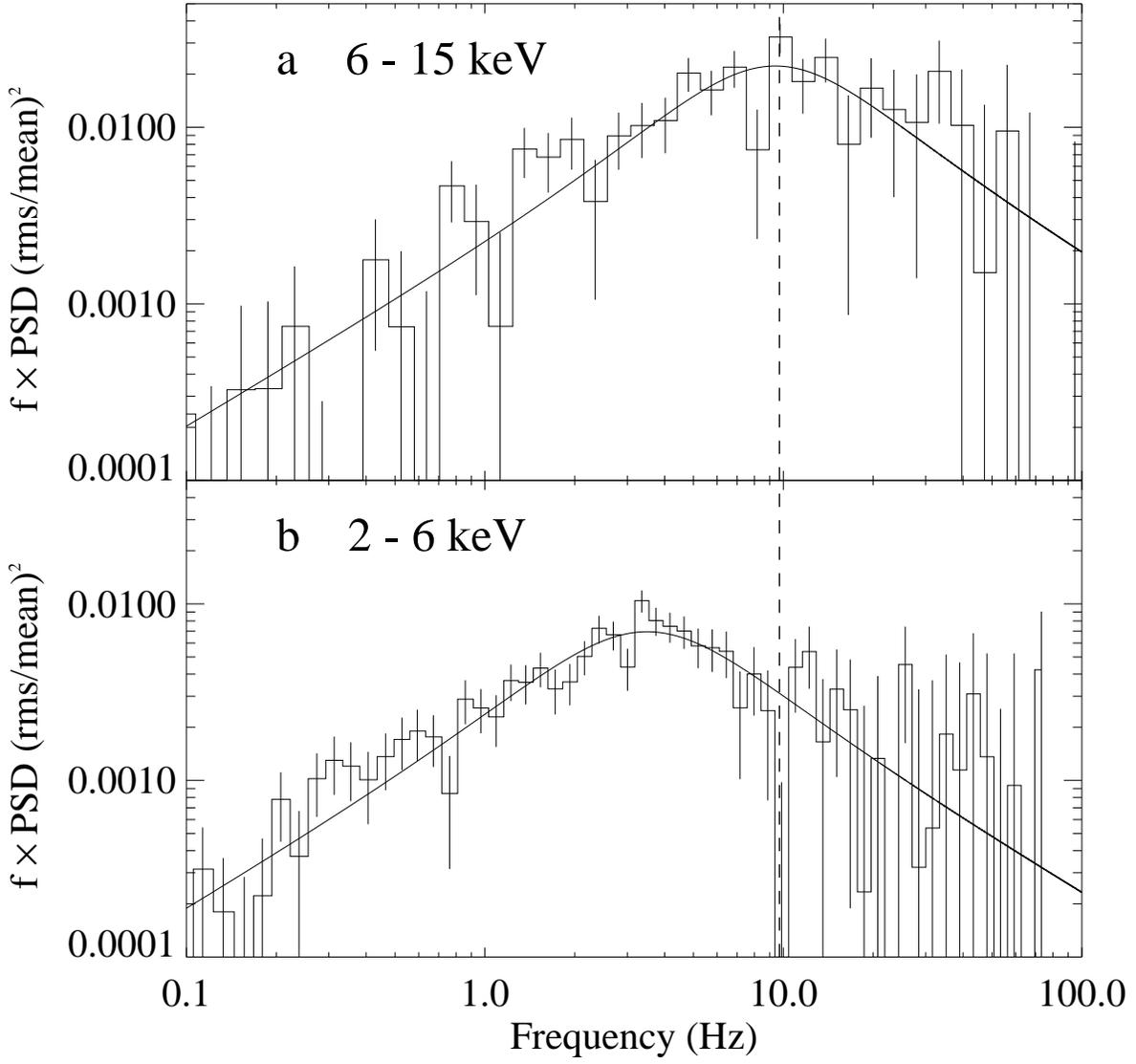}
\caption{\label{fig:datmod} 
Fits to the power spectra of Obs~1 in two energy bands: (a) 6 -- 15 keV, (b) 
2 -- 6 keV. As clearly seen, the 6 -- 15 keV Lorentzian peaks at a higher 
frequency and has higher rms amplitude for this observation. The dashed line 
is the peak frequency of the Lorentzian in the 6 -- 15 keV band. 
}
\end{figure}

We also study the dependence of temporal properties on energy. We fit the 
PSDs from two energy band light curves, 2 -- 6 keV and 6 -- 15 keV, and 
study the evolution of the peak frequencies and the rms amplitudes of 
Lorentzian components. Fig.~\ref{fig:frqsh} summarizes our efforts, where we 
plot the peak frequencies and the rms amplitudes of \La\ in the two energy 
bands mentioned. For Obs~1, during the state transition, there is a clear 
shift with energy in the peak frequency of \La , the only component in the PSD
 fit as seen in Fig.~\ref{fig:datmod}. The 6 -- 15 keV \La\ peaks at 
$3.50 \pm 0.42$ Hz, whereas the 2 -- 6 keV \La\ peaks at $9.70 \pm 2.16$ Hz. 
Moreover, the 6 -- 15 keV rms amplitude is higher than the 2 -- 6 keV 
amplitude. For the remaining observations, the 2 -- 6 keV rms amplitude is 
higher, and the Lorentzians peak at the same frequency (within 1-$\sigma$  
error).  

\begin{figure}
\plotone{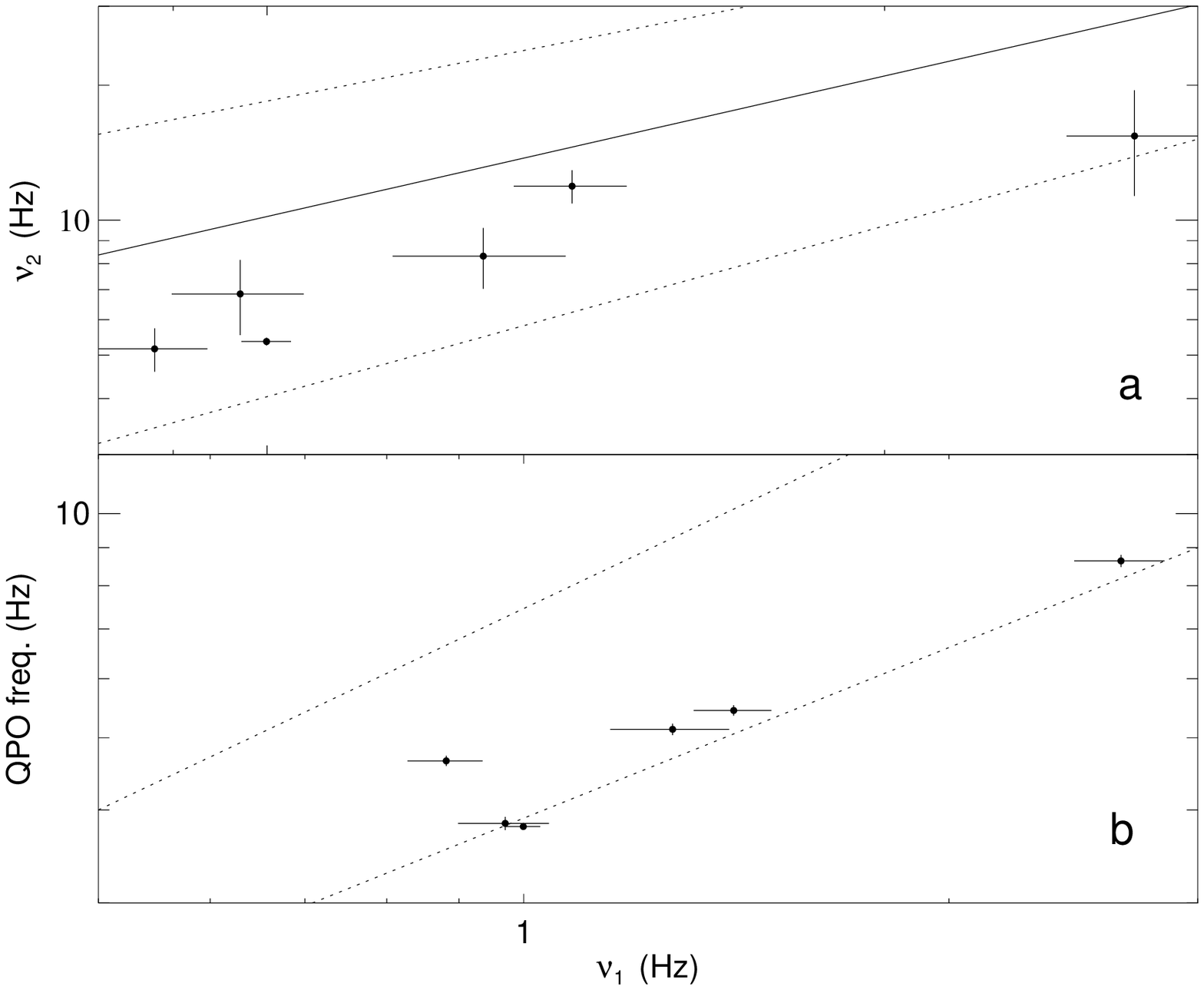}
\caption{\label{fig:psal} 
Frequency correlations: (a) Correlation between the peak frequencies of 
$L_{1}$ and $L_{2}$. The solid line is from \cite{Psaltis99} which is their 
fit to the correlation between the narrow QPO frequency and the frequency of a 
peaked Lorentzian component that occurs at frequencies higher than the QPOs. 
The dashed lines are 1$\sigma$ away from the solid line fit. (b) Correlation
between $\nu_{1}$ and the QPO frequency. The dotted lines represent the 
frequency range at which this correlation exists for other sources reported in 
\cite{Wijnands99}.
}
\end{figure}

It was shown recently that $\nu_{i}$ is an important parameter in terms 
of frequency correlations \citep{Nowak00,vanStraaten01,Pottschmidt02}.
Having observed these correlations in other sources, we decided to look for 
them in \SO , and plotted peak and QPO frequencies as a function of each
other. Two sets of parameter correlations are apparent as shown in 
Fig.~\ref{fig:psal}; $\nu_{2}$ and $\nu_{1}$ show a correlation similar to the 
one reported by \cite{Psaltis99}, and $\nu_{1}$ shows a correlation with the 
QPO frequency which is shown to exist for other black hole and neutron star 
sources \citep{Wijnands99}. Due to low statistics and large errors, we could 
not conclude whether $\nu_{3}$ correlates with $\nu_{1}$.

%

\subsection{Time lags and coherence}

   One can use the cross-spectral elements of the Fourier transform to obtain 
other information. The coherence function, for example, is a 
Fourier-frequency-dependent measure of the degree of linear correlation 
between two concurrent time series, in this case light curves measured 
simultaneously in two energy bands \citep{Nowak99}. For example, unity 
coherence means that there exists a linear transformation function between two 
energy bands for each piece of light curve that we are averaging and this 
function is the same for every piece. The Fourier time lag is a 
Fourier-frequency-dependent measure of the time delay between two concurrent 
time series \citep{Miyamoto89,Nowak99}. It is related to the phase of the 
average cross power spectrum between the ``soft energy'' light curve and the 
``hard energy'' light curve. We use the convention that the sign of the lag is 
positive when hard photons lag soft photons. Observations of hard lags in 
BHCs have often been interpreted as evidence for Compton upscattering in a hot 
electron gas \citep{Payne80}, however simple Comptonization models have 
difficulty explaining the magnitude of lags \citep{Ford99}.

%
\begin{figure}
\plotone{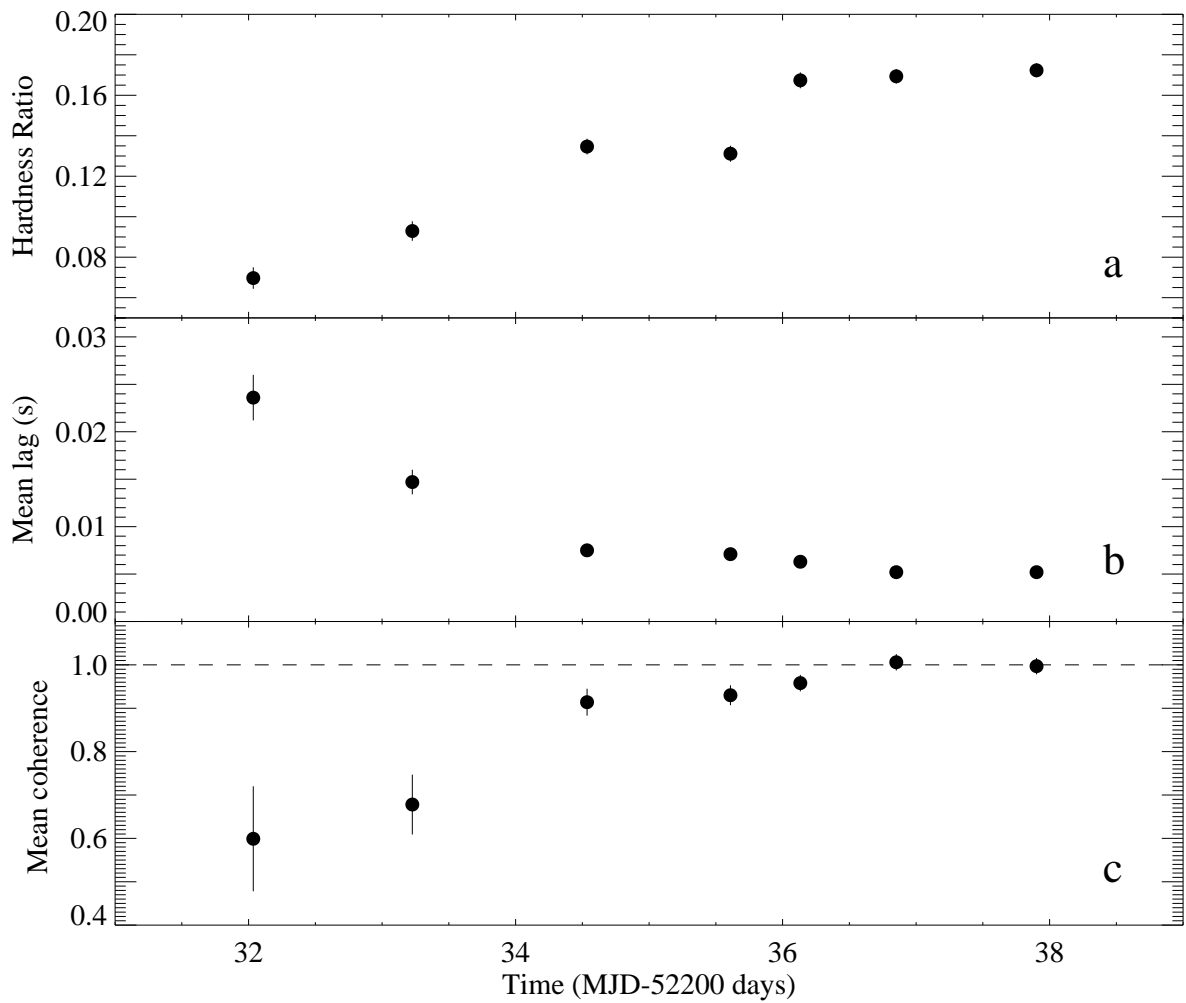}
\caption{\label{fig:coh_lags} 
(a) Ratio of the HEXTE Cluster A count rate in 17 -- 150 keV band to the PCA 
count rate per PCU in 3 -- 10 keV band (hardness ratio), (b) mean lag and 
(c) mean coherence. The mean is taken in the frequency band of 1--10 Hz where 
the measurements are well above the noise level. The energy bands are 2--6 keV
 and 6--15 keV. For Obs~1 and~2, the lags are significantly longer compared to
 the following observations. The dashed line indicates the unity coherence. 
The anti-correlation between the mean time lag and the hardness ratio, as well 
as the anti-correlation between the lags and the coherence is clearly 
observed.}
\end{figure}

    Fig.~\ref{fig:coh_lags} shows the evolution of mean coherence and mean 
time lag in 1~--~10 Hz where they are least affected by the Poisson noise. 
The mean lag decreases whereas the coherence function increases with time, 
and they both approach an asymptotic value. The time lag decays from 24 ms
to around 6 ms, and the coherence increases from 0.6 to unity. Note that during
the HS, the average coherence is consistent with being zero, therefore time 
lags have no physical meaning \citep{Nowak99}. This is the third source after 
Cyg~X-1 and GX~339$-$4 to show enhanced lags during the state transition. 
We also added the evolution of the hardness ratio of the X-ray spectrum to 
Fig.~\ref{fig:coh_lags}. The ratio is taken between the HEXTE Cluster A count 
rate in 17 -- 150 keV band and the PCA count rate per PCU in 3 -- 10 keV band. 
The hardness ratio is inversely related to the power-law index ($\Gamma$) of 
the X-ray spectrum, and preliminary spectral analysis confirms their 
anti-correlation. $\Gamma$ and average lag have been shown to correlate for 
Cyg X-1 \citep{Pottschmidt02} and GX~339$-$4 \citep{Nowak02}. The 
anti-correlation of the hardness ratio and the mean lag shows that \SO\ is 
similar to these sources.


\section{Discussion}

\subsection{Comparison of \SO\ with other sources}
  
  The temporal properties of \SO\ during the transition to the hard state are 
very similar to what has been observed for other BHCs, namely
 4U~1630$-$47 \citep{Tomsick00}, Cyg~X-1 \citep{Pottschmidt02}, and also
 XTE~J1550$-$564 \citep{Kalemci01}. For 4U~1630$-$47, a sharp transition was 
seen with most of the changes in the temporal properties occurring in less 
than 2 days during the decay of the 1998 outburst. Similarly, for \SO, 
Obs~1 and our last HS observation are only one day apart. For 4U~1630$-$47
(during its 1998 decay) and for XTE~J1550$-$564 (during its 2000 decay), the 
QPO frequency and the break frequency showed a decreasing trend, whereas the 
overall rms amplitude showed an increasing trend after the transition 
\citep{Tomsick00,Kalemci01}, as observed in \SO. On the other hand, the 3 -- 
20 keV X-ray flux decreased during and after the transition for these sources,
 whereas it increased during the transition for \SO. This is likely due to the 
hardening of the power law index and the energy range of PCA. This behavior 
can be explained if we adapt the two parameter description of \cite{Homan01} 
and assume that the second parameter is the size of the Comptonizing region. 
Then, the mass accretion rate through the disk may decrease, but the 
increasing size of the Comptonizing region may result in many hard photons in 
the PCA range which can increase the X-ray flux. We also note that there could 
be significant flux below 3 keV, and bolometric fluxes might be considerably 
higher, especially for the HS.

   Finally, it seems that the changes observed in \SO\ during the transition to
 the hard state resemble best to the transitional episodes of Cyg X-1. 
First of all, the mean X-ray time lag in 1 -- 10 Hz frequency range is longer 
and the mean coherence is low during the state transition as compared to later 
observations, just as in Cyg X-1 \citep{Cui97_2,Pottschmidt02}. The 
peak frequencies of $L_{1}$ and $L_{2}$ are significantly higher and $L_{3}$ 
is absent during the state transition, again very similar to what has been 
observed in Cyg X-1. Correlations between the peak frequencies of the 
Lorentzians, which we will discuss in Section~\ref{sec:cor}, are also similar.

\subsection{Lags and Coherence}

   During the transition, the average time lags are enhanced compared to
the following observations as shown in Fig.~\ref{fig:coh_lags}. As mentioned
earlier, this behavior has been seen before for Cyg X-1 and for GX~339$-$4
\citep{Pottschmidt00,Nowak02}. In Cyg X-1, the mean lags do not change during 
the LS and the soft state\footnote{The soft state in Cyg X-1 is probably not
the usual soft state as seen in other BHCs, but rather closer to the ``very
high state'', and therefore direct comparisons with the HS of \SO\ and soft 
state of Cyg X-1 might not be correct.} and they have similar values, but they 
only increase during the failed state transitions. \cite{Pottschmidt00} 
suggested that one possible reason is the optically thin radio outflows during 
the state transition which creates an additional Comptonizing area at its base 
that increases the time lags. It seems that the quenching of jets during the HS
 and the IS/VHS, optically thin outflows (and simultaneous building of a 
compact jet) during the state transitions, and optically thick emission during 
the LS are typical properties of BHCs \citep{Corbel00,Corbel01,Fender01b}. For 
\SO , there is no radio measurement during the state transition, but an 
inverted radio spectrum is observed during the hard state (S. Corbel, private 
communication), and it is likely that a large outflow occurred during the 
state transition as in other BHCs. Therefore, extended and hot material from
 optically thin ejections or from a compact jet building up could be the 
reason for the enhanced lags as hypothesized for Cyg X-1 and GX~339$-$4, and 
the reduced coherence could simply be a result of additional scatterings in 
the hot material. The spectral modeling of the corona and the reflection 
features also seem to support this idea for Cyg X-1 and GX~339$-$4 
\citep{Nowak02,Pottschmidt02}. 

   The reason for the lack of enhanced lags after Obs~2 could also be 
due to different Lorentzian components acting at the same frequency range. 
\cite{Nowak00} discussed the case that the overall measured time lag may be a 
composite of time lags inherent to each individual coherent PSD component. 
Although the individual time lags can be fairly large, the composite lag might 
be small. For Obs~1 the only apparent component present is $L_{1}$. Although
we can not rule out the presence of a QPO, its rms upper limit amplitude is 
low compared to rms amplitude of $L_{1}$. For Obs~2, the 1 -- 10 Hz band is 
dominated by $L_{1}$ (14\% rms amplitude), where $L_{2}$ contributes little 
(2\% rms amplitude). Moreover, the QPO, which is also within this frequency 
range, is weak. For the remaining observations, significant part of 
$L_{2}$ moves into this frequency range\footnote{For example, for Obs~5, the 
relative rms contributions of \La\ and \Lb\ in 1 -- 10 Hz range are 11\% and 
15\%, respectively.} and QPOs are strong (see Fig.~\ref{fig:lor}). This could 
be the reason for the reduced time lags after Obs~11. Note that similar 
arguments have been raised to explain the enhanced lags in Cyg~X-1 
\citep{Pottschmidt02}. One way to check this argument is to compare the 
frequency dependent lags with the shape of the power spectrum. Unfortunately, 
the quality of our time lag spectrum is not good enough to do this. However, 
the anti-correlation of the mean coherence with the mean lag seems to counter 
this argument since one would expect a decrease in the mean coherence in a 
frequency band where different processes are contributing. A steady state is 
reached after our third observation, and the time lags (correspondingly the 
phase lags) are small. For these cases, the Lorentzians could be due to 
individual coherent processes. One can still obtain near unity coherence if 
the phase lags are small as illustrated in \cite{Nowak99_2}, see especially 
equation (3) of this paper. In Fig.~\ref{fig:coh_lags}, we also included the 
evolution of the hardness ratio, since it anti-correlates with the mean time 
lag. Note that it also anti-correlates with the peak frequencies as in the 
case of Cyg X-1 and GX~339$-$4. The implications of this relation have been 
discussed extensively for GX~339$-$4 by \cite{Nowak02}.

\subsection{\label{sec:cor}Correlations between fit parameters}

   Various papers address the correlations between certain fit parameters of
the PSD for X-ray binaries and show that these correlations exist
for decades of frequency range 
\citep{Wijnands99,Psaltis99,vanStraaten01,Belloni02}. One correlation is 
between the break frequency (nearly equivalent of $\nu_{1}$) and the frequency
 of a ``bump'' or QPO above this break for different types of X-ray binaries: 
black hole candidates, atoll sources, millisecond X-ray pulsar and Z sources
\citep{Wijnands99}. We obtained a box of frequency range from Fig. 2a of their 
paper which the correlation exists and over-plotted this to our QPO frequency 
versus $\nu_{1}$ plot\footnote{We also performed power-law fits and confirmed 
that the break frequency is close to $\nu_{1}$.  When we plot the break 
frequency - QPO frequency correlation from these fits, we obtained slightly 
better agreement with the Wijnands correlation.} in Fig.~\ref{fig:psal}a. This 
plot shows that we can add \SO\ to the list of various other X-ray binaries 
for which this correlation holds. 

  There is another correlation reported by \cite{Psaltis99} between the narrow 
QPO frequency and the frequency of a peaked Lorentzian component that occurs at
 frequencies higher than the QPOs. In our case this could be the QPO frequency
vs $\nu_{2}$ or QPO frequency vs $\nu_{3}$. The former is not consistent with 
the correlation reported by \cite{Psaltis99}, and the latter is consistent 
within 1$\sigma$ error bars, however the error bars on $\nu_{3}$ are quiet 
large (see Table~\ref{table:par_lor}). The interesting result is that the 
correlation between $\nu_{1}$ and $\nu_{2}$ for \SO\ is consistent with the
 correlation reported by \cite{Psaltis99} as shown in Fig.~\ref{fig:psal}b. 
This is not very surprising since all these correlations have issues with the 
identification of the related components 
\citep{Wijnands99,Psaltis99,Belloni02}. The same correlation between $\nu_{1}$ 
and $\nu_{2}$ is seen for Cyg X-1 also, adding one more to the similarities 
list.

\subsection{\label{sec:discussion}Energy dependence of PSD components}

Perhaps the most interesting result of this paper is the shifting of the peak 
frequency of \La\ to a higher frequency with increasing energy for Obs~1 as 
shown in Fig.~\ref{fig:datmod}.  This kind of behavior has been seen with 
\emph{Ginga} in two sources before, GX~339$-$4 and GS~1124$-$68 
\citep{Belloni97}. All those observations were in the VHS, except for one 
observation in GS~1124$-$68, where it was during the transition to the LS. 
This is the second time such a shift has been observed during a state 
transition. \cite{Belloni97} interpreted the shifts as a result of shots 
with energy dependent amplitudes and profiles. We show in Fig.~\ref{fig:frqsh}
 that after Obs~1, there is no shift in the peak frequencies. This is hard to 
explain with the shot-noise models since it requires sudden removal of the 
energy dependence from the shot profile. 

We attribute the frequency shift to a geometrical effect. We adapt the idea of
the corona being the part of a large vertically stretched conical outflow 
which explains enhanced lags and reduced coherence as discussed in 
\cite{Pottschmidt02}. One can add a temperature profile to the vertical scale 
such that the base of the outflow has a higher temperature. Then, the seed 
photons from the inner disk will interact predominantly with the higher 
temperature part of the jet at its base and will be scattered to higher 
energies. Assuming that the higher frequency variability in the accretion disk 
is created close to the inner edge, the high energy power spectrum would peak 
at higher frequencies. Similarly, seed photons originating away from the inner
edge, which creates the low frequency variability, would interact predominantly
with the cooler, higher altitude parts of the outflow. When the large outflow 
is over, the corona shrinks in the vertical scale which would cause a decrease 
in the X-ray lags and an increase in the coherence. Moreover, the temperature
variation could be either absent or greatly reduced compared to the large 
outflow case which would also explain the disappearance of the frequency 
shift as observed for \SO. 

As mentioned earlier, the shift in the peak frequencies as a function of energy
has been observed during the VHS and transitions. The radio emission is 
quenched in the IS/VHS, but a corona probably exists. Although this corona is 
probably not related to the ``base of the jet'' or an ``optically thin 
outflow'' as suggested in our model to explain \SO\ during the transition, the
 overall geometry in the IS/VHS might be similar, perhaps on a smaller scale. 
A systematic search on the frequency shift effect during the VHS for \SO , and 
other sources during the IS/VHS and during the transitions can increase the 
number of constraints on the geometry of the accretion flow. 

\section{Summary and Conclusions}

The similarities of temporal properties to those of other known black 
hole systems suggest that \SO\ is also a black hole system. Especially the 
similarities to Cyg X-1 are remarkable. Here is a summary of the temporal 
properties:
\begin{itemize}
\item The state transition is sharp in terms of the temporal properties with 
the appearance of band limited noise in the PSD. The rms amplitude increases 
from 4\% upper limit in HS to $\sim$16\% during Obs~1. The last HS observation
and Obs~1 are only one day apart. 
\item The PSD can be characterized by broad Lorentzians and QPOs. The
characteristic frequencies of these components decrease after the transition 
to the LS occurs, which can be interpreted as retreating of the inner
 disk radius as discussed in \cite{diMatteo99} for other BHCs. The total rms 
amplitude also increases during this time.
\item The peak frequencies obey the correlations reported by \cite{Wijnands99}
and \cite{Psaltis99}.
\item The average lags are higher during the first two observations than the
remaining observations. Likewise, the hardness ratio shows an increasing 
behavior as the transition occurs. The coherence function anti-correlates with
mean lag and increase during the transition, and becomes unity coherence for 
the last two observations.
\item \La\ peaks at different frequencies for different energy bands during 
Obs~1. After this observation this shift disappears. Also, the rms amplitude
of \La\ increases with energy only for the first observation whereas it 
decreases with energy for the remaining observations. The shift can be a sign 
of temperature gradient at the base of the jet as discussed in
 Section~\ref{sec:discussion}.  
\end{itemize}
 

\acknowledgments 
EK acknowledges useful discussions with J\"{o}rn Wilms, Mike Nowak and Craig 
Markwardt. The authors would like to thank all scientists contributed to the 
T\"{u}bingen Timing Tools. EK was partially supported by T\"UB\.ITAK. JAT 
acknowledges partial support from NASA grant NAG5-10886. KP was supported by 
grant Sta 173/25-1 and Sta 173/25-3 of the Deutsche Forschungsgemeinschaft. 
RW was supported by NASA through Chandra Postdoctoral Fellowship grant number 
PF9-10010 awarded by CXC, which is operated by SAO for NASA under contract 
NAS8-39073. PK acknowledges partial support from NASA grant NAG5-7405.



\begin{thebibliography}{}

\bibitem[\protect\astroncite{{Augusteijn}, {Coe} \&
  {Groot}}{2001}]{Augusteijn_IAU01}
{Augusteijn}, T., {Coe}, M., \& {Groot}, P.,  2001, IAU~Circular, 7710

\bibitem[\protect\astroncite{{Belloni}, {Psaltis} \& {van der
  Klis}}{2002}]{Belloni02}
{Belloni}, T., {Psaltis}, P., \& {van der Klis}, M.,  2002, ApJ, submitted,
  astro-ph/0202213

\bibitem[\protect\astroncite{{Belloni} et~al.}{1997}]{Belloni97}
{Belloni}, T., {van der Klis}, M., {Lewin}, W.~H.~G., {van Paradijs}, J.,
  {Dotani}, T., {Mitsuda}, K., \& {Miyamoto}, S.,  1997, A\&A, 322, 857

\bibitem[\protect\astroncite{{Bradt}, {Rothschild} \& {Swank}}{1993}]{Bradt93}
{Bradt}, H.~V., {Rothschild}, R.~E., \& {Swank}, J.~H.,  1993, A\&AS, 97, 355

\bibitem[\protect\astroncite{{Castro-Tirado}
  et~al.}{2001}]{Castro_Tirado_IAU01}
{Castro-Tirado}, A.~J., {Kilmartin}, P., {Gilmore}, A., {Petterson}, O.,
  {Bond}, I., {Yock}, P., \& {Sanchez-Fernandez}, C.,  2001, IAU~Circular, 7707

\bibitem[\protect\astroncite{{Corbel} et~al.}{2000}]{Corbel00}
{Corbel}, S., {Fender}, R.~P., {Tzioumis}, A.~K., {Nowak}, M., {McIntyre}, V.,
  {Durouchoux}, P., \& {Sood}, R.,  2000, A\&A, 359, 251

\bibitem[\protect\astroncite{Corbel et~al.}{2001}]{Corbel01}
Corbel, S., et~al., 2001, ApJ, 554, 43

\bibitem[\protect\astroncite{{Cui} et~al.}{1997}]{Cui97_2}
{Cui}, W., {Zhang}, S.~N., {Focke}, W., \& {Swank}, J.~H.,  1997, ApJ, 484, 383

\bibitem[\protect\astroncite{{di Matteo} \& {Psaltis}}{1999}]{diMatteo99}
{di Matteo}, T., \& {Psaltis}, D.,  1999, ApJ, 526, L101

\bibitem[\protect\astroncite{{Fender} et~al.}{1999}]{Fender99}
{Fender}, R., et~al., 1999, ApJ, 519, L165

\bibitem[\protect\astroncite{{Fender}}{2001}]{Fender01b}
{Fender}, R.~P.,  2001, MNRAS, 322, 31

\bibitem[\protect\astroncite{{Fender} et~al.}{2001}]{Fender01}
{Fender}, R.~P., {Hjellming}, R.~M., {Tilanus}, R.~P.~J., {Pooley}, G.~G.,
  {Deane}, J.~R., {Ogley}, R.~N., \& {Spencer}, R.~E.,  2001, MNRAS, 322, L23

\bibitem[\protect\astroncite{{Ford} et~al.}{1999}]{Ford99}
{Ford}, E.~C., {van der Klis}, M., {M\'endez}, M., {van Paradjs}, J., \&
  {Kaaret}, P.,  1999, ApJ, 512, L31

\bibitem[\protect\astroncite{{Groot} et~al.}{2001}]{Groot_IAU01}
{Groot}, P., {Tingay}, S., {Udalski}, A., \& {Miller}, J.,  2001, IAU~Circular,
  7708

\bibitem[\protect\astroncite{{Homan} et~al.}{2001}]{Homan01}
{Homan}, J., {Wijnands}, R., {van der Klis}, M., {Belloni}, T., {van Paradijs},
  J., {Klein-Wolt}, M., {Fender}, R., \& {M{\'e}ndez}, M.,  2001, ApJS, 132,
  377

\bibitem[\protect\astroncite{{Kalemci} et~al.}{2001}]{Kalemci01}
{Kalemci}, E., {Tomsick}, J.~A., {Rothschild}, R.~E., {Pottschmidt}, K., \&
  {Kaaret}, P.,  2001, ApJ, 563, 239

\bibitem[\protect\astroncite{{Levine} et~al.}{1996}]{Levine96}
{Levine}, A.~M., {Bradt}, H., {Cui}, W., {Jernigan}, J.~G., {Morgan}, E.~H.,
  {Remillard}, R., {Shirey}, R.~E., \& {Smith}, D.~A.,  1996, ApJ, 469, L33

\bibitem[\protect\astroncite{{Miller} et~al.}{2002a}]{Miller02}
{Miller}, J.~M., {Fabian}, A.~C., {Wijnands}, R., {Reynolds}, C.~S., \& {Ehle},
  M.,  2002a, ApJ, 570, 69

\bibitem[\protect\astroncite{{Miller} et~al.}{2002b}]{Miller_ATEL02}
{Miller}, J.~M., {Wijnands}, R., {Wojdowski}, P., {Groot}, P., {Fabian}, A.~C.,
  {van der Klis}, M., \& {Lewin}, W.,  2002b, The Astronomer's Telegram, \#81

\bibitem[\protect\astroncite{{Miyamoto} \& {Kitamoto}}{1989}]{Miyamoto89}
{Miyamoto}, S., \& {Kitamoto}, S.,  1989, Nature, 342, 773

\bibitem[\protect\astroncite{{Nowak}}{2000}]{Nowak00}
{Nowak}, M.~A.,  2000, MNRAS, 318, 361

\bibitem[\protect\astroncite{{Nowak} et~al.}{1999}]{Nowak99}
{Nowak}, M.~A., {Vaughan}, B.~A., {Wilms}, J.~., {Dove}, J.~B., \& {Begelman},
  M.~C.,  1999, ApJ, 510, 874

\bibitem[\protect\astroncite{{Nowak}, {Wilms} \& {Dove}}{1999}]{Nowak99_2}
{Nowak}, M.~A., {Wilms}, J.~., \& {Dove}, J.~B.,  1999, ApJ, 517, 355

\bibitem[\protect\astroncite{{Nowak}, {Wilms} \& {Dove}}{2002}]{Nowak02}
{Nowak}, M.~A., {Wilms}, J., \& {Dove}, J.~B.,  2002, MNRAS, submitted,
  astro-ph/0201383

\bibitem[\protect\astroncite{Payne}{1980}]{Payne80}
Payne, D.,  1980, ApJ, 237, 951

\bibitem[\protect\astroncite{{Pottschmidt}}{2002}]{Pottschmidt02th}
{Pottschmidt}, K.,  2002, PhD Thesis, Univ. of T{\"u}bingen

\bibitem[\protect\astroncite{{Pottschmidt} et~al.}{2000}]{Pottschmidt00}
{Pottschmidt}, K., {Wilms}, J., {Nowak}, M.~A., {Heindl}, W.~A., {Smith},
  D.~M., \& {Staubert}, R.,  2000, A\&A, 357, L17

\bibitem[\protect\astroncite{{Pottschmidt} et~al.}{2002}]{Pottschmidt02}
{Pottschmidt}, K., et~al., 2002, A\&A, submitted, astro-ph/0202258

\bibitem[\protect\astroncite{{Psaltis}, {Belloni} \& {van der
  Klis}}{1999}]{Psaltis99}
{Psaltis}, D., {Belloni}, T., \& {van der Klis}, M.,  1999, ApJ, 520, 262

\bibitem[\protect\astroncite{{Remillard}}{2001}]{Remillard_IAU01}
{Remillard}, R.,  2001, IAU~Circular, 7707

\bibitem[\protect\astroncite{{Tanaka} \& {Lewin}}{1995}]{Tanaka95}
{Tanaka}, Y., \& {Lewin}, W. H.~G.,  1995,
\newblock in {X}-ray {B}inaries, ed. W.~H.~G. {Lewin}, J. {Van Paradjs},
  E.~P.~J. {Van den Heuvel},  (Cambridge: Cambridge {U}. {P}ress),  126

\bibitem[\protect\astroncite{{Tomsick} et~al.}{2002a}]{Tomsick_inprep02}
{Tomsick}, J.~A., et~al., 2002a, in preperation

\bibitem[\protect\astroncite{{Tomsick} \& {Kaaret}}{2000}]{Tomsick00}
{Tomsick}, J.~A., \& {Kaaret}, P.,  2000, ApJ, 537, 448

\bibitem[\protect\astroncite{{Tomsick} et~al.}{2002b}]{Tomsick_iau02}
{Tomsick}, J.~A., {Kalemci}, E., {Corbel}, S., \& {Kaaret}, P.,  2002b,
  IAU~Circular, 7837

\bibitem[\protect\astroncite{{van Straaten} et~al.}{2001}]{vanStraaten01}
{van Straaten}, S., {van der Klis}, M., {di Salvo}, T., {Belloni}, T., \&
  {Psaltis}, D.,  2001, ApJ, in press, astro-ph/0107562

\bibitem[\protect\astroncite{{Wijnands}, {Miller} \&
  {Lewin}}{2001}]{Wijnands_IAU01}
{Wijnands}, R., {Miller}, J.~M., \& {Lewin}, W. H.~G.,  2001, IAU~Circular,
  7715

\bibitem[\protect\astroncite{{Wijnands} \& {van der Klis}}{1999}]{Wijnands99}
{Wijnands}, R., \& {van der Klis}, M.,  1999, ApJ, 514, 939

\bibitem[\protect\astroncite{Zhang et~al.}{1995}]{Zhang95}
Zhang, W., Jahoda, K., Swank, J.~H., Morgan, E.~H., \& Giles, A.~B.,  1995,
  ApJ, 449, 930

\end{thebibliography}


\begin{table}[ht]
\caption{\label{table:par_lor} Lorentzian Fit Parameters (a)}
\begin{minipage}{\linewidth}
\scriptsize
\begin{tabular}{c|c|c|c|c|c|c|c} \hline \hline
Obs.  & 1 & 2 & 3  & 4 & 5 & 6 & 7 \\ 
Date (MJD)\footnote{MJD at the beginning of the observation.}  & 52232.0 & 52233.2  & 52234.5  & 52235.1  & 52236.1  & 52236.9  & 52237.9 \\ \hline
Flux\footnote{Flux in 2--20 keV range, in units of $\rm 10^{-9}\,ergs\,cm^{-2}\,s^{-1}$.} & 2.60 & 2.75 & 2.68 & 2.56 & 2.30 & 2.29 & 2.15 \\
Int. time\footnote{Integration time in seconds.} & 1224 & 1488 & 1152 & 2000 & 1664 & 1824 & 1904 \\
HR $\times\,10^{3}$ \footnote{Ratio of the HEXTE Cluster A count rate in 17 -- 150 keV band to the PCA count rate per PCU in 3 -- 10 keV band, multiplied by
$10^{3}$.}  & $\rm 70\pm5$  & $\rm 93\pm5$ & $\rm 135\pm4$ & $\rm 131\pm4$ & $\rm 167\pm4$ & $\rm 169\pm4$ & $\rm 172\pm3$ 
\\ \hline
$\nu_{1}$ & $\rm 4.18\pm0.51$ & $\rm 2.65\pm0.19$ & $\rm 1.28\pm0.12$ & $\rm 1.41\pm0.09$ & $\rm 0.88\pm0.05$ & $\rm 1.00\pm0.03$ & $\rm 0.97\pm0.07$ \\
$L_{1}$ rms & $\rm 17.28\pm0.85$ & $\rm 18.94\pm0.90$ & $\rm 17.72\pm1.26$ & $\rm 21.28\pm0.56$ & $\rm 18.71\pm0.50$ & $\rm 22.81\pm0.20$ & $\rm 24.00\pm0.82$ \\
$L_{1}$ fwhm & $\rm 7.96\pm0.79$ & $\rm 3.81\pm0.37$ & $\rm 1.88\pm0.24$ & $\rm 2.34\pm0.15$ & $\rm 1.18\pm0.07$ & $\rm 1.71\pm0.04$ & $\rm 1.65\pm0.12$ \\ \hline
$\nu_{2}$ & - & $\rm 15.41\pm4.08$ & $\rm 8.31\pm1.29$ & $\rm 11.91\pm1.01$ & $\rm 5.16\pm0.57$ & $\rm 4.68\pm0.08$ &  $\rm 6.85\pm1.30$  \\
$L_{2}$ rms & - & $\rm 12.79\pm2.27$ & $\rm 20.26\pm2.07$ & $\rm 9.30\pm1.44$ &$\rm 26.21\pm0.80$ & $\rm 15.97\pm1.94$ & $\rm 16.38\pm2.59$ \\
$L_{2}$ fwhm & - & $\rm 19.06\pm5.14$ & $\rm 14.10\pm1.41$ & $\rm 8.11\pm1.98$ & $\rm 11.42\pm0.71$ & $\rm 5.23\pm0.17$  & $\rm 8.85\pm1.07$  \\ \hline
$\nu_{3}$ & - & - & - & $\rm 80.49\pm40.71$ & $\rm 85.70\pm30.60$ & $\rm 25.49\pm2.77$ & $\rm 84.99\pm37.25$ \\
$L_{2}$ rms & - & - & - & $12.87\pm3.02$ & $15.07\pm1.40$ & $\rm 16.17\pm0.52$ & $\rm 12.64\pm2.68$ \\
$L_{3}$ fwhm & - & - & - & $151.56\pm58.87$ & $171.41\pm62.20$ & $\rm 50.97\pm5.50$ &  $\rm 131.67\pm54.74$ \\ \hline
QPO freq.\footnote{This is the resonance frequency, not the peak frequency} & 0.1-20\footnote{Frequency range searched.} & $\rm 8.64\pm0.16$ & $\rm 5.13\pm0.09$ & $\rm 5.44\pm0.09$ & $\rm 4.66\pm0.07$ & $\rm 3.80\pm0.03$ & $\rm 3.84\pm0.08$ \\
QPO rms & $<$2.8\footnote{95\% confidence upper limit.} & $\rm 3.98\pm1.06$ & $\rm 7.10\pm1.12$ & $\rm 13.12\pm0.80$ & $\rm 4.36\pm0.94$ & $\rm 3.72\pm0.27$ & $\rm 8.83\pm2.09$ \\
QPO fwhm & 0.167 $\times$ freq.\footnote{The width is fixed to 0.167 $\times$ QPO frequency ($Q$ = 6).} & $\rm 0.98\pm0.56$ & $\rm 1.34\pm0.39$ & $\rm 3.29\pm0.38$ & $\rm 0.62\pm0.27$ & $\rm 0.39\pm0.08$ & $\rm 1.57\pm0.48$ \\ \hline
$\rm \chi^{2}/\nu$ & 56.0/67 & 144.1/144 & 153.6/144 & 166.8/152 & 125.5/141 & 182.7/141 & 137.6/141 \\ \hline
\end{tabular}
\end{minipage}
\end{table}


\end{document}